\documentclass[11pt,superscriptaddress,floatfix,preprint]{revtex4-1}
\usepackage{graphicx}
\usepackage{bm,amsmath,amssymb,mathrsfs,dcolumn}
\usepackage{subfigure}
\usepackage{units}
\usepackage{hyperref}
\usepackage[usenames]{color}
\hypersetup{pdfborder=0 0 0,colorlinks=true,citecolor=blue,linkcolor=blue}

\begin{document}

\title{Role of Polar Phonons in the Photo Excited State of Metal Halide 
Perovskites}
\date{\today}

\author{Menno Bokdam}
\email{menno.bokdam@univie.ac.at}
\author{Tobias Sander}
\affiliation{Faculty of Physics, Computational Materials Physics, University of 
Vienna, Sensengasse 8/12, 1090 Vienna, Austria}

\author{Alessandro Stroppa}
\author{Silvia Picozzi}
\affiliation{Consiglio Nazionale delle Ricerche - CNR-SPIN, I-67100
L'Aquila, Italy}

\author{D.D. Sarma}
\affiliation{Solid State and Structural Chemistry Unit, Indian Institute of 
Science, 560012 Bangalore, India}

\author{Cesare Franchini}
\author{Georg Kresse}
\affiliation{Faculty of Physics, Computational Materials Physics, University of 
Vienna, Sensengasse 8/12, 1090 Vienna, Austria}

\begin{abstract}
The development of high efficiency perovskite solar cells has sparked a 
multitude of measurements on the  optical properties of these materials. For 
the most studied methylammonium(MA)PbI$_3$ perovskite, a large range 
(6-55 meV) of exciton binding energies has been reported by various 
experiments. 
The existence of 
excitons at room temperature is unclear. For the MAPb$X_3$ 
perovskites we report on relativistic 
Bethe-Salpeter Equation calculations ($GW$-BSE). This method is 
capable to directly calculate 
excitonic properties from first-principles. At low 
temperatures it predicts exciton binding energies in agreement with the 
reported 'large' values. For MAPbI$_3$, phonon modes present in this frequency 
range have a negligible contribution to the ionic screening. By calculating the 
polarization in time from finite temperature molecular dynamics, we 
show that at room temperature this does not change. We therefore exclude 
ionic screening as an explanation for the experimentally observed reduction of 
the exciton binding energy at room temperature and argue in favor of the 
formation
of polarons.
\end{abstract}

\maketitle

\hyphenation{organo-metal}

%%%%%%%%10%%%%%%%%20%%%%%%%%30%%%%%%%%40%%%%%%%%50%%%%%%%%60%%%%%%%%70%%%%%%%%80

In the last three years metal halide perovskites have come up as very 
promising solar cell 
materials\cite{Green:natpt14,Graetzel:natm14,Kim:jpcc14}. 
Because of their 
relatively simple production 
procedure and high photovoltaic efficiency, they bear the 
potential of becoming competitive 
with current silicon based solar cells. The materials have an $OMX_3$ 
perovskite 
structure (Organic($O$), Metal($M$), Halide($X$)) and 
depending on the temperature up to three different crystal 
phases. The most frequently studied material is MAPbI$_3$. At temperatures 
above 333~K, the lead and iodine atoms form a cubic perovskite structure  
enclosing a methylammonium (MA) molecule\cite{Stoumpos:ic13}.
Combinations with the halogens Cl and Br can also be made and result 
in perovskite structures with different volumes and larger band gaps, not 
ideally suited for solar applications. According 
to early experimental measurements from the 1990s and 2000s, the MAPb$X_3$ 
perovskites are semiconductors with optical band gaps ($\Delta_{\rm 
opt}$) ranging between $\sim$1.6-3.1 
eV\cite{Hirasawa:jpsj94,Papavassiliou:sm95,Tanaka:ssc03,
Kitazawa:joms02}. The optical gap is slightly lower than the fundamental 
electronic band gap ($\Delta$), because of 
the 
electron-hole (e-h) interaction present in the excited system. With the 
emergence of very efficient perovskite solar cells, 
the mechanism behind the material's  
good energy conversion rate  has become a focus of research. In 
this regard, one important issue  is the relatively large exciton binding 
energies ($\rm{E_{xb}}$) 
reported for these materials, 6-55 meV for MAPbI$_3$
\cite{Hirasawa:jpsj94,Tanaka:ssc03,Yamada:ieeep15,Miyata:natp15,Sun:ees14,
Saba:natc14,Savenije:jpcl14,Zhang:nanol14,Innocenzo:natc14}
 and 76 meV\cite{Tanaka:ssc03} for MAPbBr$_3$.  Intriguingly, many reported 
values are higher than $k_{\rm B}T$, which should make it difficult for 
electrons and holes to separate after excitation. It is then a mystery
why these materials are so efficient in converting solar energy to power. As a 
solution
to this puzzle, it has been proposed that 
ionic contributions from the PbI$_3$ framework and the
rotational freedom of MA molecules contribute to the screening 
properties, 
thereby reducing the exciton binding 
energy\cite{Huang:prb13,Frost:nanol14,Even:jpcc14,Menendez:arxiv15}. Alternative 
explanations
invoke the formation of polarons, quasiparticles dressed by the ionic lattice 
that might lower
the band gap\cite{Menendez:arxiv15} below the excitonic onset.
Furthermore, recent experiments indicate that temperature 
also plays a role. Y.Yamada $et$ 
$al.$\cite{Yamada:ieeep15} measured a reduction of $\rm E_{xb}$ from $\sim$30 
meV at 13 K to $\sim$6 meV at 300 K, and likewise, A.Miyata $et$ 
$al.$\cite{Miyata:natp15}. 
measured a value of 16$\pm 2$ meV in the low temperature orthorhombic 
phase, but only a few meV at room temperature. Whether ionic screening does or 
does not affect $\rm E_{\rm xb}$ is under debate\cite{Filippetti:jpcl15}. The 
large range of the 
reported $\rm E_{xb}$ values indicate the  need for a theoretical description. 
In this work we report about first 
principles calculations on these ionic systems and address excitons in their 
interplay with polar phonons. 

\begin{figure}
\includegraphics[scale=0.7]{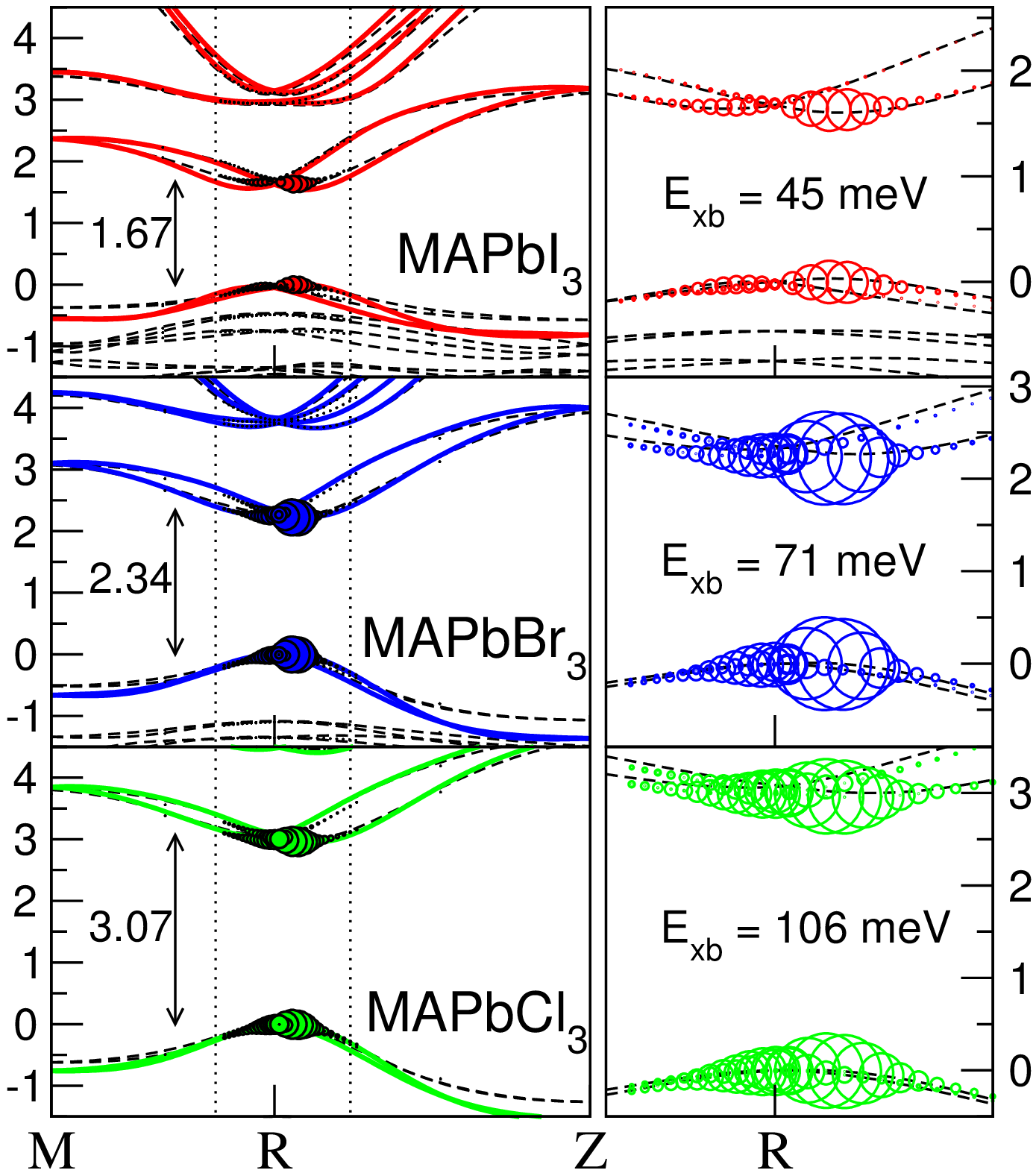}
\caption{Left: $GW_0$ band structure of MAPb$X_3$ in the 
pseudo-cubic phase, with $X=$ I (red), Br (blue) and Cl (green)
determined by Wannier interpolation\cite{Mostofi:cpc08} from a calculation 
using 
$4\times 4\times 4$ k-points. The band 
gaps at $R$ are indicated in eV. The dashed 
lines in the background are the corresponding DFT+scissor band structures. 
Right: Zoom-in of the band structure (marked by the dotted lines) close to the 
$R$ point. The radii of the circles represent the contribution of the e-h pair 
at that k-point ($|A^{1}_{c,v,\mathbf{k}}|$) to the first exciton 
wave function.} 
\label{fig1}
\end{figure}

%%%%%%%%10%%%%%%%%20%%%%%%%%30%%%%%%%%40%%%%%%%%50%%%%%%%%60%%%%%%%%70%%%%%%%%80

The theoretical modeling of metal halide 
perovskites is extremely challenging as it involves the treatment of several 
subtle, but important effects that are difficult to compute accurately. 
The first issue is the lattice structure to consider. 
Temperature dependent crystal structures have been 
determined\cite{Stoumpos:ic13}, but uncertainties in the orientation of 
the organic part prevent an unequivocal structural resolution. The structural 
characteristics have been the subject of numerous studies 
based on different (Local Density, Generalized Gradient, van der Waals) DFT 
approximations\cite{Borriello:prb08,Mosconi:jpcc13,Brivio:aplm13,Feng:jpcl14,
Egger:jpcl14,Motta:natc15}. For the cubic phase of MAPbI$_3$, most 
calculations predict lattice constants in good to excellent agreement with 
experiment. However, differences in the orientation of the 
molecule and the resulting deformation of the unit cell have been reported. 
We address this issue here
(i) by performing global search for minimum energy structures, (ii) calculating 
excitons for various unit cells, (iii) and finite 
temperature simulations. Second, the presence of heavy elements requires to 
consider  
relativistic effects including spin-orbit coupling. 
\cite{Filip:prb14,Proupin:prb14} 
Furthermore, for a quantitative description of the electronic 
structure,  it is essential to calculate many-body quasiparticle energies
e.g. in the framework of the $GW$ 
approximation\cite{Umari:sr14,Brivio:prb14,Filip:prb14,Castelli:aplm14}. 
Finally, to evaluate $\rm{E_{xb}}$ and calculate accurate optical spectra, it 
is 
necessary to account for the e-h interaction. This 
can be done by the Bethe-Salpeter equation (BSE)  following
$GW$ calculations\cite{Castelli:aplm14,Even:pssrrl13}--- a computationally
exceedingly challenging endeavor  if spin-orbit coupling is taken into account. 
It is therefore not
astonishing that previous work has often  given unsatisfactory results.
Several quasiparticle $GW$ calculations have been reported recently, 
\cite{Umari:sr14,Brivio:prb14,Filip:prb14,Castelli:aplm14}, but a 
fully relativistic treatment including spin-orbit interaction was only performed 
in Refs. 
\cite{Filip:prb14,Ahmed:epl14}. Although, BSE calculations have been reported,
these often neglect relativistic effects\cite{Even:pssrrl13,Castelli:aplm14} and 
report
much too large binding energies.  Even if relativistic effects are accounted 
for,
the binding energies  ($\rm{E_{xb}}=0.153$ eV) are at least a {\em factor 3} too 
large compared 
to any experimental values\cite{Ahmed:epl14}. As we will show here, we can 
entirely resolve this issue when sampling the Brillouin zone
with sufficient accuracy.

\textit{\color{red}Computational method.}
The first-principles calculations use a 
plane-wave basis and the 
projector augmented wave (PAW) 
method\cite{Blochl:prb94b} as implemented in the {\sc vasp} 
code\cite{Kresse:prb93,Kresse:prb96,Kresse:prb99}. For structure determination, 
the 
PBEsol (Perdew, Burke, 
Ernzerhof modified for solids)\cite{Perdew:prl08} functional  was used, if  not 
otherwise noted.
Cross checks were also performed using van der Waals corrected functionals, 
specifically,  
the PBE-D3  method of Grimme\cite{Grimme:jcp10} finding no relevant differences 
for the 
properties reported here.
%PBEsol yields volumes in excellent agreement with experiment. Although van der 
%Waals (vdW) interactions are not explicitly included, the PBEsol functional 
%yields weak physisorption minima at about the correct distance for vdW bonded 
%systems. 
The  MAPb$X_3$ cubic perovskite unit cells (12 atoms per cell) were 
constructed
starting from the cubic-phase of MAPbI$_3$ determined by 
X-ray diffraction\cite{Stoumpos:ic13} and seeking the global energy minimum by 
simulated annealing. To determine candidate structures, 
molecular
dynamics simulations were performed with a linear decrease of the temperature 
from 800~K to 500~K in 50000 steps of 1.5~fs. Approximately every $\sim$1000 
steps a snapshot was taken and fully relaxed. This process was 
repeated from the lowest energy structure yet found. A unique global minimum 
was found for
all considered materials.(See Supplementary Materials) In the subsequent  
electronic structure calculations ($GW$ and BSE), SOC  was 
fully included, 
and for Pb the  $5s^25p^65d^{10}$ orbitals were included in 
the valence\cite{Filip:prb14}. Gaussian smearing with 
$\sigma=0.05$~eV was used to broaden the one-electron levels. Many-body 
effects 
were accounted for by first calculating  PBE orbitals, and then 
determining  the quasiparticle energies and fundamental gaps in the $GW_0$ 
approximation\cite{Hedin:pr65,Hybertsen:prb86}.  Here  the one electron 
energies in $G$ were iterated until the quasiparticle 
energies are converged, while keeping $W_0$ fixed at the DFT-RPA 
level\cite{Shishkin:prb06}. About 2100 empty bands on a $4\times4\times4$ 
$\Gamma$-centered k-point grid and 128 points on the frequency grid are needed 
to obtain well converged band gaps. 

To determine the optical properties, the Bethe-Salpeter equation for the 
polarizability\cite{Hanke:prb80,Onida:rmp02,Sander:prb15} was solved. The 
common 
 Tamm-Dancoff approximation\cite{Dancoff:prb50}, 32 occupied and unoccupied KS 
orbitals, the $W_0$ of
the preceding $GW_0$ calculations,  and 
$6\times6\times6$ k-points centered on a low symmetry k-point were used.  

To obtain k-point converged values for the exciton binding energy $\rm{E_{xb}}$ 
at least $20\times20\times20$ k-points are, however, 
required. These BSE calculations, were performed  using only 2 (un)occupied 
orbitals and  fitting $W_0$ to a model dielectric  
function\cite{Bechstedt:ssc92} that depends parametrically on the 
macroscopic dielectric constant determined in the 
previous BSE calculations with few k-points. Since even $GW$ calculations are 
prohibitive for so many
k-points, we use PBE calculations and applied a scissor technique to raise 
the unoccupied KS eigenvalues (compare Fig. 1). At these dense k-point 
grids, the $\rm{E_{xb}}$ becomes linearly dependent on the inverse of the total 
number of k-points\cite{Fuchs:prb08}. The $\rm{E_{xb}}$ values reported in this 
work are therefore obtained by linear extrapolation to obtain the limit of 
the infinitely dense k-point grid.(See Supplementary Materials)

The effect of 
different molecular orientations on the exciton binding 
energy have been assessed by BSE calculations on low energy 
configurations of 
the $\sqrt{2}\times\sqrt{2}$ FASnI$_3$ and MAPbI$_3$ super cells. In 
addition, we have constructed a 
$\sqrt{2}\times\sqrt{2}\times 2$ super cell for FASnI$_3$. These structures 
were acquired by taking snap shots from Parallel Tempering Molecular Dynamics 
(PTMD) calculations at 300 K. The 
$\sqrt{2}\times\sqrt{2}$ structures are the lowest energy 
configurations from the PTMD trajectory and were 
relaxed into their instantaneous ground state, while keeping the volume and 
cell shape fixed to the experiment. The 
$\sqrt{2}\times\sqrt{2}\times 2$ structure is a randomly picked configuration 
at 300 K from a separate PTMD calculation and was not relaxed. In the 
$\sqrt{2}\times\sqrt{2}$ 
structures the molecular dipoles are orthogonally orientated w.r.t. each other 
and in the 
$\sqrt{2}\times\sqrt{2}\times 2$ structures all the molecular dipoles have a 
different 
orientation. The same BSE calculation procedure was used as before, but the 
screening parameters and $GW_0$ gap were not calculated; the values for the 
unit cell were used instead. This is a reasonable approximation, since 
calculation of the
screening in the computationally more efficient random phase approximation 
shows 
little
difference in electronic screening for different unit cells and different 
molecular
orientations.

Details of the finite temperature dielectric function calculations are presented 
in the results section.

\begin{table}[!b]
\caption{Effective electron and hole masses ($m_{e/h}^*$) of the VBM and CBM in 
the MR and RZ directions from the $GW_0+$SOC band structures, the ratio of 
the exciton effective mass ($\mu$) over the high freq. dielectric constant 
squared
($\varepsilon^2_{\infty}$), the Wannier-Mott -($\rm E^{WM}_{xb}$) the BSE 
calculated ($\rm E_{xb}$) exciton binding energies.}
\label{table:par}
\begin{ruledtabular}
\begin{tabular}{lccccccc}
$X_3$& MR & ZR & MR & ZR & $\mu/\varepsilon^2_{\infty}$& $\rm 
E^{WM}_{xb}$& 
$\rm E_{xb}$\\ \hline
 &($m_e^*$)& ($m_e^*$)&($m_h^*$)& ($m_h^*$)& & (meV)& (meV)\\
 \hline
I&0.19&0.17&0.28&0.23&$\nicefrac{0.11}{6.83^2}$ &32& 45\\
 \hline
Br&0.26&0.22&0.35&0.25&$\nicefrac{0.13}{5.15^2}$ &67& 71\\
 \hline
Cl&0.39&0.30&0.38&0.32&$\nicefrac{0.18}{4.22^2}$ &138& 106\\
\end{tabular}
\end{ruledtabular}
\end{table}

\textit{\color{red}Results.} 
In Figure \ref{fig1} (left), the calculated $GW_0$ quasiparticle band 
structures of 
the three MAPb$X_3$ structures are shown. The band gap at the $R$ 
points 
is indicated and is in excellent agreement with 
experiment\cite{Hirasawa:jpsj94,Papavassiliou:sm95,Tanaka:ssc03,
Kitazawa:joms02}. SOC shifts the band gap minimum to $R'$ making it slightly 
indirect\cite{Brivio:prb14}. The exciton wave function is expressed in an 
electron-hole 
product basis, 
$\Phi^1=\sum_{cv\mathbf{k}}A^{1}_{c,v,\mathbf{k}}\phi_{c,\mathbf{k}}\phi_{v, 
\mathbf{k}}$. The 
first eigenstate $\{A^{1}_{c,v,\mathbf{k}}\}$ of the generalized BSE eigenvalue 
problem\cite{Sander:prb15} is visualized by plotting $|A^{1}_{c,v,\mathbf{k}}|$ 
as a fat band structure.  On the right hand side of Fig. 
\ref{fig1} a zoom-in of the region close to $R$ is made. It shows that the 
exciton is very localized in k-space, primarily consisting of states at the 
band extrema. Going from iodine to chlorine, the dispersion flattens (effective 
electron/hole masses increase), the band gap increases and, as a result, the 
extent of the exciton in k-space increases. The corresponding parameters are 
tabulated in Table \ref{table:par}. We have calculated the corresponding 
exciton binding energies also
in the Wannier-Mott (WM) model  for screened Coulomb interacting e-h pairs in 
parabolic bands: $\rm E^{WM}_{xb}=(\mu/\varepsilon^2_{\infty})\rm{R}_{\infty}$, 
with $\mu^{-1}={\overline{m_e}^*}^{-1}+{\overline{m_h}^*}^{-1}$ 
the effective mass of the e-h pair, $\varepsilon_{\infty}$ the high 
freq. dielectric constant and $\rm{R}_{\infty}$ the Rydberg 
constant. Since we use the SOC split
"Rashba-Dresselhaus"\cite{Stroppa:natc14} band structure in the BSE method, we 
can test the validity of the simple parabolic dispersion assumed in the WM 
model. We see (Tab.\ref{table:par}) that WM gives the correct order of the e-h 
interaction, however it results in a different ratio between I:Br and Br:Cl, 
which can not be trivially explained by small errors in $\mu$ or 
$\varepsilon_{\infty}$.

\begin{figure}[!t]
\includegraphics[scale=0.45]{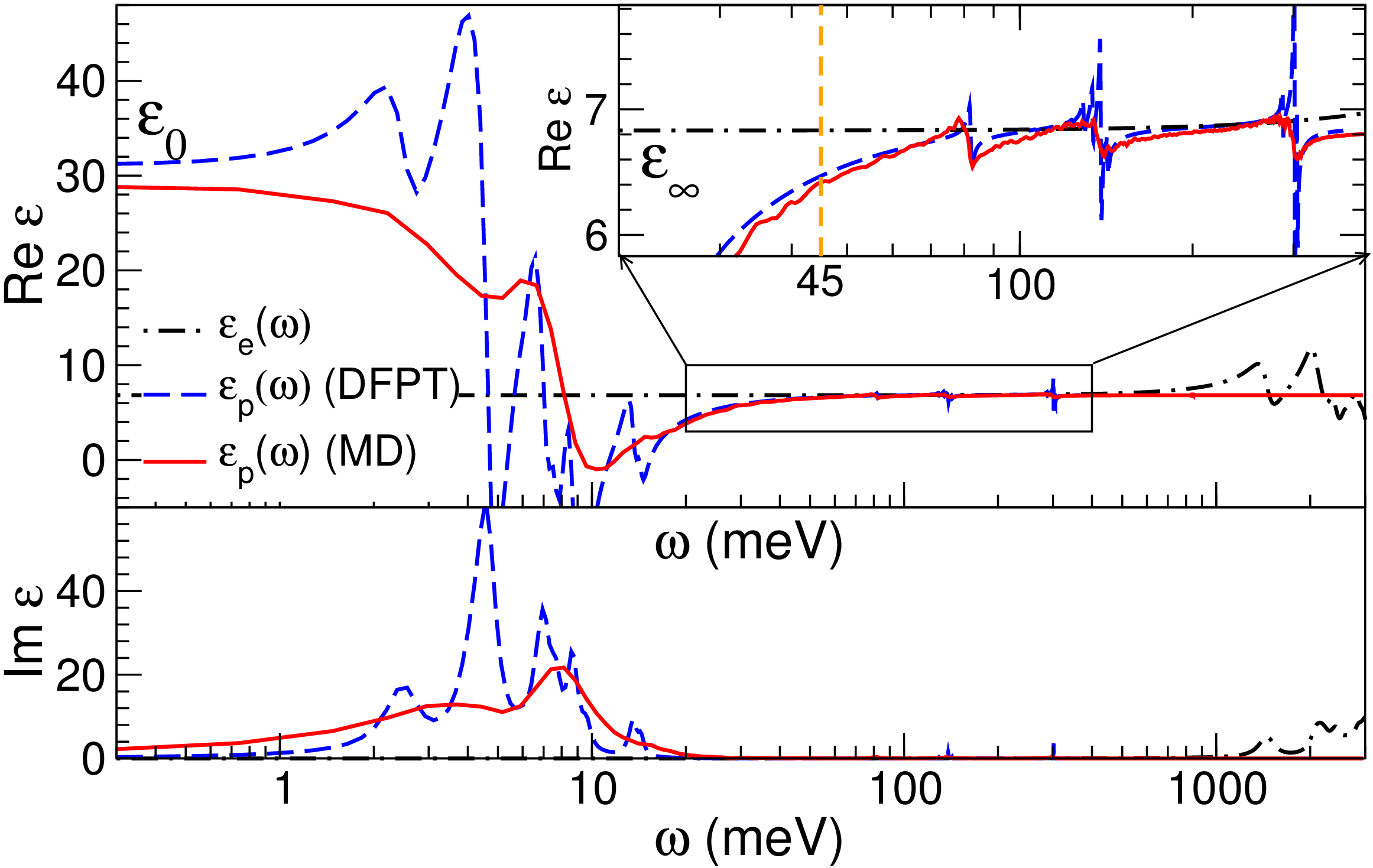}
\caption{Ionic and electronic contributions ($\varepsilon_{\rm 
p}$, 
$\varepsilon_{\rm e}$) to the dielectric 
function $\varepsilon({\omega})$ of MAPbI$_3$. The 
top/bottom figure shows the real/imaginary part of $\varepsilon({\omega})$. The 
solid red/dashed blue line are, respectively the results from the 
DFPT(T=0K)/MD(T=300K) 
method. The inset shows a zoom-in of $\rm Re\,\varepsilon_{\rm p}(\omega)$ 
close to the $\rm E_{\rm xb}$ of 45 meV.} 
\label{fig3}
\end{figure}

%%%%%%%%10%%%%%%%%20%%%%%%%%30%%%%%%%%40%%%%%%%%50%%%%%%%%60%%%%%%%%70%%%%%%%%80
An important question is, whether the ionic contributions to the screening
can be disregarded in the BSE calculations. To explore this point,  Figure 
\ref{fig3} shows the
sum of the ionic and electronic  contribution to the dielectric function at 0 K, 
$\varepsilon(\omega)$, with the
ionic contribution calculated using density functional perturbation theory 
(DFPT)\cite{Wu:prb05,Gajdos:prb06,Perez:jpcc15}.
A sizable increase of the  static dielectric constant ($\varepsilon_0$) 
compared to  the 'ion-clamped' high frequency dielectric constant 
($\varepsilon_{\infty{}}$) is found. The increase comes from optically active 
phonon modes below 20 meV (see Im $\varepsilon_{\rm p}(\omega)$), clearly 
displaying the ionic nature of this material. However, the phonon modes present 
in the 
relevant energy 
window around $\rm E_{\rm xb} \approx 45$ meV (see inset) are practically not 
active.
 
Since the exciton binding seems to change with temperature, the second 
intriguing question is whether the screening  changes at finite temperature.
To explore this, we have developed a novel scheme to evaluate the
dielectric ionic response at finite temperature that we briefly describe in this 
section
(a more detailed description will be presented in a future work).
The idea is inspired by methods usually used to determine the {\em electronic
contributions} to the screening in time-dependent 
DFT\cite{Walter:jcp08}. 
Well equilibrated finite temperature ensembles are subjected to a short 
constant electric field in time 
$E\delta(t)$ acting on the {\em ions}. The 
$\delta$-pulse is a 
natural way of exciting all possible frequencies in the system. The force 
exerted by this field onto the ions is proportional to
${\rm F}_{\alpha} = \sum_{\beta} {\rm Z}_{\alpha,\beta} {\rm E}_{\beta},$
where ${\rm Z}_{\alpha,\beta}$ are the Born effective charges evaluated by 
density
functional perturbation theory, and $\alpha$ and $\beta$ are Cartesian
indices\cite{Wu:prb05}. In the first 
time step, these forces are added, thereby exciting the ionic system. The issue 
is to find a suitable way to calculate the induced ionic
polarization ${\rm P}(t)$ caused by the delta peak. Here, we  calculate the
induced polarization as $\delta {\rm P}(t)=({\rm P}_{+}(t)-{\rm 
P}_{-}(t))/2, $ where ${\rm P}_{+}(t)$ is the time evolution of the  
polarization for a positive delta peak $E\delta(t)$, and
${\rm P}_{-}(t)$ the time evolution of the polarization after a negative delta 
peak $-E\delta(t)$.  The evolving ${\rm P}(t)$ can in principle be 
evaluated using the Berry curvature\cite{Resta:rmp94},
but the Berry curvature often jumps discontinuously as the ions move.
Hence, we evaluate the change of the polarization from the velocities
$v_{\alpha}(t)$ and the Born effective charges ${\rm Z_{\alpha\beta}}(t)$ as
${\rm 
P}_{\beta}(t) = \int_0^t \frac{d{\rm 
P}_{\beta}}{dt'} dt'= \int_0^t \frac{d{\rm 
P}_{\beta}}{du_{\alpha}}\frac{du_{\alpha}}{dt'} dt'=
   \int_0^t {\rm Z_{\beta\alpha}}(t') \cdot v_{\alpha}(t') dt'.$
The additional cost is small, since $\mathbf{\rm Z}(t)$ varies very
slowly and needs to be recalculated only about every 50 time steps. 
 The Fourier transformation of $\delta P(t)$ is 
directly related to the ionic polarizability\cite{Thomas:pccp13}. We first 
tested
this approach at $T=0~$K and found exact agreement with perturbation theory.
To obtain
reasonably noise-free data at finite temperature, we use a $2\times2\times2$ 
super cell and average
over 80 starting configurations in order to converge the 
spectrum. After the $\delta$-pulse, the system is allowed to 
evolve in the  micro-canonical ensemble  
unperturbed for 3~ps, the short time somewhat limiting the spectral 
resolution. 
However, the 80 starting configurations were obtained by a 
taking independent snapshots every 0.7~ps from a well equilibrated 60~ps long 
finite 
temperature MD trajectory. Therefore, we expect most of the dynamics to be 
sampled. The PBE-D3  method of 
Grimme\cite{Grimme:jcp10} was used here (although PBEsol 
results are very similar) and the deuterium mass was used for 
the hydrogen atoms. This replacement only changes the hydrogen related modes
above 100 meV and allows to increase the time step during the simulation.

Clearly, the 300 K finite temperature polarizability (solid red line in 
Fig.\ref{fig3}) above 20 
meV is very similar to the one at T=0~K (dashed blue line in 
Fig.\ref{fig3}). 
The modes are at 
the same positions but broadened by  fluctuations in the cage structure, as well 
as rotations 
of the molecules. Below 20 meV some differences are visible, however,  in both 
methods the calculated 
$\varepsilon_0$ is close 
to 30,  in excellent agreement with the measured value of
28.8\cite{Poglitsch:jcp87}.

We now discuss the question whether ionic screening should be included
in the calculation of the exciton binding energy. The values reported in Tab.  
\ref{table:par}
assume a fixed lattice,  i.e. a vertical transition. It is well 
established from the Franck-Condon
energy diagram that lattice relaxation in the excited state 
can only {\em decrease} the transition energy, i.e.  thermal or adiabatic 
transition energies including relaxation 
are always below vertical transition energies\cite{Freysoldt:rmp14}. Lattice 
relaxation therefore can
only increase the exciton binding energy, since the fundamental gap minus
the transition energy is  defined as the exciton binding energy ${\rm E_{\rm 
xb}}$.
Whether relaxation needs to be included, i.e. whether the vertical or thermal 
transition energy
is measured experimentally can be disputed, although common wisdom is that
optical absorption virtually always measures vertical excitation energies. 
One argument is that  
the position of the (dominant) longitudinal optical phonon mode ($\omega_{\rm 
LO}$) compared to the typical energy scales of the optical absorption determines
whether ionic relaxation should be considered. In the 
effective mass approximation  this  leads to two extreme cases, 
$\rm E_{\rm xb}\gg\hbar\omega_{\rm LO}$, $\varepsilon_{\rm 
eff}\rightarrow\varepsilon_\infty$ and 
$\rm E_{\rm 
xb}\ll\hbar\omega_{\rm LO}$ (ionic relaxation needs to be 
included)\cite{Bechstedt:prb05,Bechstedt:book15}. 
Since the dominant 
active phonon modes for MAPbI$_3$ are all below 10 meV, we have $\rm E_{\rm 
xb}\approx5\hbar\omega_{\rm LO}$ and the use of an $\varepsilon_{\rm eff} 
\approx \varepsilon_\infty \approx 6.8$ 
 is entirely justified (see Fig. \ref{fig3} for energies larger than $40$~meV). 
The temperature independence of the ionic screening, furthermore, implies that 
the observed lowering of 
the exciton binding energy at elevated temperatures must have a different 
origin than changes in the ionic screening.
In agreement with theory, recent room temperature
time-resolved terahertz spectroscopy experiments, indicate a near constant 
screening $\varepsilon(\omega)=5.5$ in the frequency range of 
$\omega=$40-100 meV and $\rm E_{\rm xb}=49\pm3$ meV\cite{Valverde:ees15}. 

However, if the exciton binding energy $\rm E_{\rm xb}$ is not lowered by ionic 
screening, 
what mechanism then leads to carrier separation at higher temperatures? Our 
calculations also shed light on
this. Individual electrons  $e^-$ and holes $h^+$ can be screened by the 
lattice, thereby forming 'dressed' quasiparticles (QP) known as polarons. Since 
the mesoscopic
Wannier-Mott model was so precise, we again resort to a mesosocopic model, 
namely,
Fr\"ohlich's theory for large polarons.  In this model, polaron formation lowers 
the QP energy by  $\rm 
E_{p}=(-\alpha-0.0123\alpha^2)\hbar\omega_{LO}$, with a coupling constant 
$\alpha 
= (\frac{1}{\varepsilon_{\infty}}-\frac{1}{\varepsilon_{0}})
(\frac{\mathrm{e^2}}{\hbar})(\frac{m^*}{2\hbar\omega_{\rm LO}})^{1/2}$. Using 
the data from Table \ref{table:par}, a screening of  $\varepsilon_0=30, 
\varepsilon_{\infty}=6$, and 
$\hbar\omega_{\rm LO}=8$ meV from Fig. \ref{fig3} we obtain an $\alpha$ of 
2.3/2.8.
This lowers the QP energy of the electron and hole by 19 and 23 meV, 
respectively,
and hence reduces the QP gap by 42 meV. This means that the charge separated 
polaronic state is only
slightly less stable than the bound exciton. If we further recall that
after excitation the electrons and holes are not yet close to
the conduction or valence band edges, and that they are both individually 
scattered by lattice phonons
loosing energy but possibly gaining momentum\cite{Bernardi:prl14}, it is likely 
that they will
rapidly separate in space and never reach their global groundstate, the bound 
exciton.
Charge separation after optical excitation will be further eased 
by non-regularities in the electrostatic potential. And 
non-regularities exist aplenty in MAPbI$_3$ at elevated temperatures:
the polar MA molecules seem to prefer a short range  ferroelectric order causing 
ferroelectric domains 
and a strong corrugation of the electrostatic potential\cite{Frost:nanol14}. 
A possible way to experimentally disentangle polaron formation and such 
molecular contributions
and related corrugations in the potential
is to perform  control measurements on an $O$PbI$_3$ perovskite with  $O$ 
cations that are non polar, 
for instance Cs.

\begin{figure}[!t]
\includegraphics[scale=0.45]{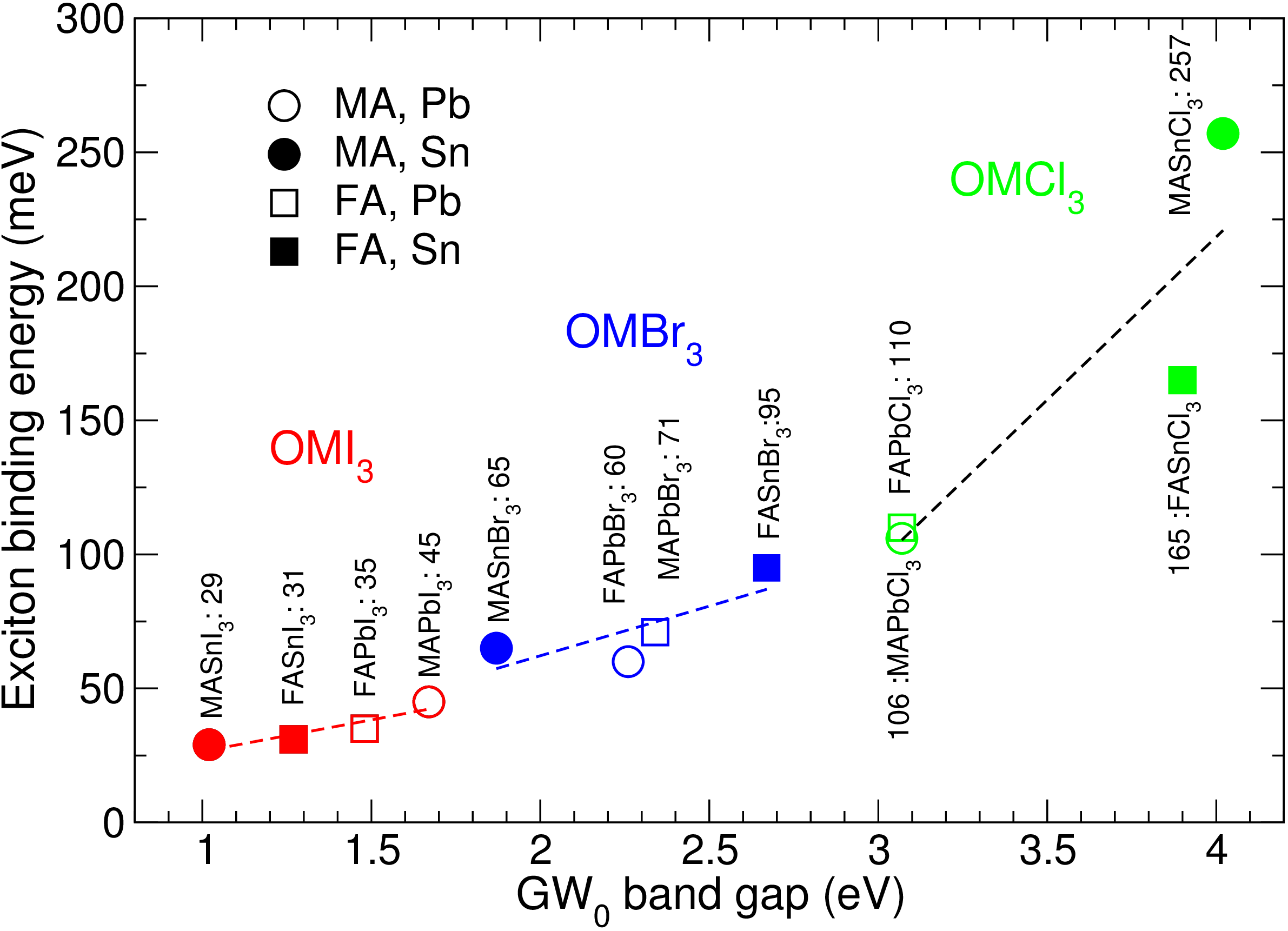}
\caption{Calculated exciton binding energies and $GW_0$ band 
gaps of twelve metal halide perovskites ($OMX_3$, \{$O={\rm MA,FA}, 
M={\rm Pb,Sn}, X={\rm I,Br,Cl}$\}).}
\label{fig5}
\end{figure}

We like to comment briefly 
on the performance of a wider class of perovskites ($OMX_3$). Specifically, we 
have 
replaced MA by formamidinium (FA) and Pb by Sn, thereby constructing twelve 
different perovskites. Their ionic lowest energy structure was 
calculated as before by 
simulated annealing and subsequent relaxation. FA is larger than MA and thereby 
changes 
the band gap of the perovskite. FASnI$_3$ is a particular interesting 
candidate, previous work suggests that this is possibly a ferroelectric 
 lead-free alternative for MAPbI$_3$\cite{Stroppa:natc14}. In Figure 
\ref{fig5} we show the trend in the exciton binding energies w.r.t. the $GW_0$ 
band gap. Clearly, the 
halogen species predominantly determines the gap. For
each halogen, the strength of the exciton binding energy and the 
optical gap can be fine-tuned by varying the molecule or the metal atom.
Nevertheless, only iodine based perovskites seem to posses sufficiently
small band gaps and exciton binding energies to be suitable for solar
cells. An overview of the band gaps calculated at the various 
level of theory and available experimental data has been presented in Table 
\ref{table1}. Over the whole range a good 
agreement is found between the $GW$-BSE calculations on these small unit cell 
structures and experimentally observed band gaps. Small discrepancies can be 
caused by the unit cell approach taken in this work. For those structures, 
which 
have not yet been synthesized or for which the band gap has not yet been 
measured, we put these number forward as predictions.

%%%%%%%%10%%%%%%%%20%%%%%%%%30%%%%%%%%40%%%%%%%%50%%%%%%%%60%%%%%%%%70%%%%%%%%80

\begin{table}
\caption{ Calculated onset of optical absorption ($\Delta_{\rm 
opt}=\Delta_{\rm GW}-\rm{E_{xb}}$), exciton 
binding 
energy ($\rm{E_{xb}}$), the GW$_0$ ($\Delta_{\rm GW}$), DFT ($\Delta_{\rm 
DFT}$) and DFT without SOC ($\Delta_{\rm 
DFT}^{\rm ws}$) band gaps. Available 
experimental results are shown for 
comparison.}
\resizebox{\columnwidth}{!}{%
\label{table1}
%\begin{ruledtabular}
\begin{tabular}{l|ccccccc}
 \hline
&$\Delta_{\rm opt}$& $\Delta^{\rm EXP}_{\rm opt}$&$\rm{E_{xb}}$&$E^{\rm 
EXP}_{\rm xb}$&$\Delta_{\rm GW}$&$\Delta_{\rm 
DFT}$&$\Delta_{\rm 
DFT}^{\rm ws}$\\
&(eV)&(eV)&(meV)&(meV)&(eV)&(eV)&(eV)
\\
 \hline
MASnI$_3$&1.00&1.21\cite{Stoumpos:ic13},1.63\cite{
Papavassiliou:sm95} &29&-&1.03&0.41&0.67\\
FASnI$_3$&1.23&1.41\cite{Stoumpos:ic13}
&31&-&1.27&0.47&0.80\\
FAPbI$_3$&1.45&1.43,\cite{Pang:com14}1.45,\cite{
Stoumpos:ic13}1.48\cite{Eperon:ees14} 
&35&-&1.48&0.56&1.55\\

MAPbI$_3$&1.63&1.52\cite{Stoumpos:ic13}
1.57,\cite{Eperon:ees14}
&45&
6\cite{Yamada:ieeep15},
16\cite{Miyata:natp15},
19\cite{Sun:ees14},
25\cite{Saba:natc14},
35\cite{Savenije:jpcl14},
&1.67&0.77&1.69\\
 & &
1.63\cite{Hirasawa:jpsj94,Tanaka:ssc03}
1.65,\cite{Papavassiliou:sm95}
& &
38\cite{Hirasawa:jpsj94},
45\cite{Zhang:nanol14},
50\cite{Tanaka:ssc03},
55\cite{Innocenzo:natc14} 
& & & \\

 \hline  
 
MASnBr$_3$&1.80&2.25\cite{Papavassiliou:sm95}
&65&-&1.87&0.84&1.06\\
FAPbBr$_3$&2.20&2.23\cite{Eperon:ees14}
&60&-&2.26&0.97&1.96\\
MAPbBr$_3$&2.27&2.26,\cite{Tanaka:ssc03}2.33,\cite{
Papavassiliou:sm95}2.35\cite{
Kitazawa:joms02} &71&76\cite{Tanaka:ssc03} 
&2.34&1.07&2.03\\
FASnBr$_3$&2.58&-&95&-&2.67&1.47&1.61\\

 \hline
 
FAPbCl$_3$&2.96&-&110&-&3.07&1.50&2.33\\
MAPbCl$_3$&2.97&3.11,\cite{Kitazawa:joms02}3.13\cite{
Papavassiliou:sm95}
&106&-&3.07&1.51&2.47\\
FASnCl$_3$&3.74&-&165&-&3.90&2.16&2.25\\
MASnCl$_3$&3.77&-&257&-&4.02&2.30&2.46\\
 \hline
\end{tabular}
}
%\end{ruledtabular}
\end{table}
%%%%%%%%10%%%%%%%%20%%%%%%%%30%%%%%%%%40%%%%%%%%50%%%%%%%%60%%%%%%%%70%%%%%%%%80

\textit{\color{red} Discussion.} 
The last point we need to consider is how different molecular orientations 
influence the results.
In a real (super)structure 
the exciton wave 
function will span many unit 
cells with molecules that have different orientations. It has been suggested 
that the ordering and orientation of 
the molecules in the lattice aids the e-h dissociation 
process \cite{Even:jpcc14,Frost:nanol14,Ma:nanol14}. Both the MA and FA 
molecules have an 
intrinsic dipole moment and are only weakly bonded to the $MX_3$ cage. It is 
known from Nuclear Magnetic Resonance Spectroscopy 
measurements that the MA molecules in MAPb$X_3$ have the full 
rotational degree of freedom at room temperature and that reorientation is a 
fairly rapid 
process\cite{Wasylishen:ssc85}. However, with a typical reorientation time in 
the pico-second time scale, it is the slowest screening mechanism present in 
the $OMX_3$ perovskites. To asses the effect of different molecular 
orientations, BSE calculations have been performed on 
$\sqrt{2}\times\sqrt{2}$ and $\sqrt{2}\times\sqrt{2}\times2$ cells containing 2 
and 4 molecules, respectively. For the $\sqrt{2}\times\sqrt{2}$ and the 
larger FASnI$_3$ cell, the calculated exciton binding energies are 33 and 31 
meV, respectively. Compared to the 31 meV predicted for the unit cell, the 
super cell approach does not give significantly different results for the 
exciton binding energy. The same holds for MAPbI$_3$, where the 
$\sqrt{2}\times\sqrt{2}$ cell results 
in an exciton binding energy of 51 meV, which is only slightly larger than the 
45 meV predicted for the unit cell.

\textit{\color{red} Conclusion.} 
Accurate first principles calculations predict  exciton 
binding energies of the order of 50, 70 and 110 meV 
for MAPbI$_3$, MAPbBr$_3$ and MAPbCl$_3$, respectively.  The agreement
of the Wannier-Mott model with our high level  calculations
is good, provided that the model parameters are taken from 
 accurate first principles calculations.
The large exciton binding energy is clearly
at variance with the observed high efficiency of solar cells, but in excellent
agreement with most low temperature measurements. The much discussed ionic 
screening is almost
temperature independent and substantially 
increases $\varepsilon_0$ from around 6 to 30. However,  the 
optically active modes are too slow  ($<10$ meV) to effectively
screen the excitons. For certain, we can rule out that a change of the  ionic 
screening is responsible for the experimentally observed reduction of the 
exciton binding energy at room temperature. Instead,  our calculations predict 
a different scenario: electrons and holes separate after optical excitation 
forming
two individual polarons, lowering the fundamental gap by 45 meV. 
This scenario should now be carefully evaluated by experiments, and   if 
validated, offers an 
intriguing option for the  design of novel {\em  polaronic} solar cell 
materials.

\acknowledgments
MB, CF and DDS acknowledge funding by the Austrian Science 
Fund (FWF) within F2506-N17 and by the joint Austrian Science Fund - Indian 
Department of Science and Technology (DST) project
INDOX (Grant No. I1490-N19). GK and CF acknowledge funding by Austrian Science 
Fund (FWF): F41 SFB ViCoM. The calculations 
were partly performed at the Vienna Scientific Cluster (VSC-3).

{\bf Author Contributions}
G.K., C.F., M.B and A.S. conceived the project;
M.B. and G.K. proposed and implemented the new method to calculate the
finite temperature dielectric function, and performed all DFT and GW-BSE
calculations. D.D. and S.P. included insight from the experimental and
theoretical standing of the perovskite field. All the authors contributed
with discussions and writing of the manuscript.

{\bf Additional information} 
Supplementary information is available in the online version of the paper. 
Reprints and permissions information is available online at 
www.nature.com/reprints.

{\bf  Competing financial interests} 
The authors declare no competing financial interests.
\bibliographystyle{naturemag}

\end{document}